\documentclass[aps,preprint]{revtex4}
\usepackage{amsfonts}
\usepackage{amsmath}
\usepackage{amssymb}
\usepackage{graphicx}

\input{tcilatex}
\begin{document}

\preprint{}
\title{A Cosmological Model without Singularity and Dark Matter}
\author{Shi-Hao Chen}
\affiliation{Institute of Theoretical Physics, Northeast Normal University, Changchun
130024, China.\\
shchen@nenu.edu.cn}

\begin{abstract}
According to the cosmological model without singularity, there are s-matter
and v-matter which are symmetric and have oppose gravitational masses. In
V-breaking s-matter is similar to dark energy to cause expansion of the
universe with an acceleration now, and v-matter is composed of v-F-matter
and v-W-matter which are symmetric and have the same gravitational masses
and forms the world$.$ The ratio of s-matter to v-matter is changeable.
Based on the cosmological model, we confirm that big bang nucleosynthesis is
not spoiled by that the average energy density of W-matter (mirror matter)
is equal to that of F-matter (ordinary matter). According to the present
model, there are three sorts of dark matter which are v-W-baryon matter
(4/27), unknown v-F-matter (9.5/27) and v-W-matter (9.5/27)$.$ Given
v-F-baryon matter (4/27) and v-W-baryon matter can cluster and respectively
form the visible galaxies and dark galaxies. Unknown v-F-matter and
v-W-matter cannot cluster to form any celestial body, loosely distribute in
space, are equivalent to cold dark matter, and their compositions are
unknown. The number in a bracket is the ratio of the density of a sort of
matter to total density of v-matter. The decisive predict is that there are
dark celestial bodies and dark galaxies. The energy of F-matter can
transform into the energy of W-matter by such a process in which the
reaction energy is high enough.
\end{abstract}

\date{}
\received[Received text]{}
\revised[Revised text]{}
\accepted[Accepted text]{}
\published[Published text]{}
\startpage{1}
\endpage{46}
\maketitle
\tableofcontents

\section{Introduction}

Mirror matter as dark matter has been presented$^{[1]}.$ In order that the
model is consistent with the standard cosmological framework, the model
ascribes the macroscopic asymmetry of the universe to asymmetric initial
conditions of matter and mirror matter$^{[2]}$. However, when temperature is
high enough, the expectation values of all Higgs fields are zero so that the
static masses of all particles are zero, and all particles can transform
from one into other.\ Consequently the energy densities of two sorts of
matter must be equal if both are symmetric so that their degrees of freedom
are equal to each other. Thus the asymmetric initial conditions make one to
be uncomfortable.

A quantum field theory without divergence has been constructed which can
obtain all results of the given quantum field theory and in which there is
no divergence of loop-corrections and the energy of the vacuum must be zero
without normal product$^{[3]}$. There must be two sorts of matter in this
quantum field theory. The two sorts of matter are called $F-matter$ and $%
W-matter$, respectively. $F-matter$ and $W-matter$ are symmetric, and there
are only gravitation and very weak interaction by Higgs bosons in low energy
between both. $F-matter$ forms the given world and $W-matter$ is dark matter
for a $F-observer^{[4]}$. This model of dark matter is equivalent to the
model of mirror matter as dark matter.

A cosmological model without singularity has been constructed$^{[5]}.$ The
model can well explain evolution of the universe, formation of large scale
structure and the features of huge voids, naturally determines the
cosmological constant to be zero, has no singularity and gives some
predictions. Based on the cosmological model, we will see in the present
paper that the average density of $W-matter$ may be equal to that of $%
F-matter$. The average density of dark matter is the $23/4$ times of that of
visible matter, because dark matter is composed of $W-matter$ and a part of $%
F-matter$.

Second 2 discusses that the average density of W-matter is equal to that of
F-matter. Second 3 discusses the sorts and average density of dark matter.
Second 4 the interaction of F-scalar fields and W-scalar fields. Section 5
is the conclusions.

\section{The average density of W-matter is equal to that of F-matter}

According to the cosmological model without singularity, the space evolving
equations are$^{[5]}$

\begin{equation}
\overset{\cdot }{R}^{2}(t)+K=\eta \left[ \rho _{v}+V_{v}(\varpi
_{v})+V_{0}-\rho _{s}-V_{s}(\varpi _{s})\right] R^{2}\left( t\right) =\eta
\left( \rho _{g}+V_{g}\right) R^{2}\left( t\right) ,  \tag{1}
\end{equation}%
\begin{eqnarray}
\overset{\cdot \cdot }{R}(t) &=&-\frac{1}{2}\eta \left[ \left( \rho
_{v}+3p_{v}\right) -2\left( V_{v}(\varpi _{v})+V_{0}\right) -\left( \rho
_{s}+3p_{s}\right) +2V_{s}(\varpi _{s})\right] R\left( t\right)  \notag \\
&=&-\frac{1}{2}\eta \left( \rho _{g}+3p_{g}-2V_{g}\right) R\left( t\right) ,
\TCItag{2}
\end{eqnarray}%
where $\omega =\Omega ,$ $\Phi ,$ and $\chi ,$ $\eta \equiv 8\pi G/3;$ $\rho
_{v}$ and $\rho _{s}$ are the mass densities of $v-matter$ and $s-matter $ $%
\left( \text{here }c=1\right) $, respectively; $V_{v}$ and $V_{s}$ are the
densities of $v-Higgs$ potential energy and $v-Higgs$ potential energy,
respectively. the curvature factor $K$ is a function of the gravitational
mass density, and 
\begin{equation}
\rho _{g}\equiv \rho _{v}-\rho _{s},\text{ \ }p_{g}=p_{v}-p_{s},\text{ \ }%
V_{g}=V_{v}(\varpi _{v})+V_{0}-V_{s}(\varpi _{s}).  \tag{3}
\end{equation}%
\begin{equation}
1\geq K>0\text{ \ for }\rho _{g}>0,\text{ }K=0\text{ \ for }\rho _{g}=0,%
\text{ \ }0>K\geq -1\text{ \ for }\rho _{g}<0.  \tag{4}
\end{equation}%
In $V-breaking$ in which the expectation values of $v-Higgs$ fields $\langle
\omega _{v}\rangle \equiv \varpi _{v}=0$, and the expectation values of $%
s-Higgs$ fields $\langle \omega _{s}\rangle ^{\prime }\equiv \varpi
_{s}^{\prime }\neq 0,$ the gravitational mass of $v-matter$ is positive and
the gravitational mass of $s-matter$ is negative.

When temperature rises to the critical temperature $T_{cr}$ because space
contracts$^{[5]},$ $\varpi _{v}=\varpi _{s}=0,$ $\rho _{v}=\rho _{s},$ $%
V_{v}(\varpi _{v})=V_{s}(\varpi _{s})=0$ so that 
\begin{eqnarray}
\overset{\cdot }{R}^{2}(t) &=&-K+\eta V_{0}R^{2}\left( t\right) ,  \TCItag{5}
\\
\overset{\cdot \cdot }{R}(t) &=&-\eta V_{0}R\left( t\right) <0.  \TCItag{6}
\end{eqnarray}%
There is the highest temperature $T_{\max }$ at which $\overset{\cdot }{R}%
=0, $ $R\left( t\right) =R_{\min }\equiv \sqrt{K/\eta V_{0}}$ and space
inflation must occur.

$V-matter$ is composed of $v-F-matter$ and $v-W-matter$, $s-matter$ is
composed of $s-F-matter$ and $s-W-matter$. $F-matter$ and $W-matter$ are
symmetric so that divergence of loop corrections is eliminated and the
energy of the vacuum is determined to be zero. There is only the gravitation
and the interaction by Higgs bosons between $F-particles$ and $W-particles.$
The interaction by Higgs bosons\ can be ignored when temperature is low
because the masses of the Higgs bosons are all very large. Consequently $%
W-matter$ is regarded dark matter relative to $F-matter$, and vice versa.
These are the necessary inferences of the quantum field theory without
divergence$^{[3]}.$ $F-matter$ and $W-matter$ correspond to ordinary matter
and mirror matter in Ref$^{[1,2]},$ respectively$.$

The state with $T\geq T_{cr}$ is such a state with the highest symmetry. $%
\rho _{sF}=\rho _{sW}=\rho _{vF}=\rho _{vW}$ and $%
T_{sF}=T_{sW}=T_{vF}=T_{vW}=T$ in this state because $F-matter$ and $%
W-matter $ are symmetric and can transform from one into another, $s-matter$
and $v-matter$ are symmetric and can transform from one into another, and
the thermal equilibrium can realized due to $\overset{\cdot }{R}\sim 0$ when 
$T\geq T_{cr}.$

After space inflation, reheating process occurs. After reheating process,
the potential energy density transforms into $\rho _{v}^{\prime }=xV_{0}$
and $\rho _{s}^{\prime }=\left( 1-x\right) V_{0}.$ $\rho _{v}^{\prime }>\rho
_{s}^{\prime }$ because $\langle \omega _{v}\rangle =0\longrightarrow
\langle \omega _{v}\rangle _{0}\neq 0,$ $\langle \omega _{s}\rangle
=0\longrightarrow \langle \omega _{s}\rangle _{0}=0,$ $V_{v}=0%
\longrightarrow -V_{0}$ and $V_{s}=0\longrightarrow 0^{[5]}.$ Thus, after
reheating process, the evolving equations become 
\begin{equation}
\overset{\cdot }{R}^{2}(t)+K=\eta \left[ \rho _{vF}+\rho _{vW}-\rho
_{sF}-\rho _{sW}\right] R^{2}\left( t\right) =\eta \left[ 2\rho _{vF}-2\rho
_{sF}\right] R^{2}\left( t\right) ,  \tag{7}
\end{equation}%
\begin{equation}
\overset{\cdot \cdot }{R}(t)=-\eta \left[ \left( \rho _{vF}+3p_{vF}\right)
-\left( \rho _{sF}+3p_{sF}\right) \right] R\left( t\right) ,  \tag{8}
\end{equation}%
Here we still denote $\rho _{v}^{\prime }+\rho _{v}$ and $\rho _{s}^{\prime
}+\rho _{s}$ by $\rho _{v}$ and $\rho _{s},$ respectively, for convenience.

It is seen from $\left( 4\right) $ and $\left( 7\right) $ that in contrast
with the conventional cosmological models, although $\rho _{vF}=\rho _{vW},$
the $\overset{\cdot }{R}^{2}(t)+K$ cannot doubled because of $\rho _{s}=\rho
_{sF}+\rho _{sW}.$ Consequently the big bang nucleosynthesis cannot be
spoiled by $\rho _{vF}=\rho _{vW}$ provided $\rho _{s}$ is suitably chosen.
For example, if $\rho _{s}=\rho _{vW}$ in the stage of the big bang
nucleosynthesis, the evolving equations $\left( 7\right) -\left( 8\right) $
become%
\begin{equation*}
\overset{\cdot }{R}^{2}(t)+K=\eta \rho _{vF}R^{2}\left( t\right)
\end{equation*}%
\begin{equation*}
\overset{\cdot \cdot }{R}(t)=-\frac{1}{2}\eta \left( \rho
_{vF}+3p_{vF}\right) R\left( t\right) ,
\end{equation*}%
This is consistent with the conventional theory.

\section{The sorts and average density of dark matter}

Recent astronomical observations show that the universe expanded with a
deceleration early and is expanding with an acceleration now. This implies
that there is dark energy$^{[6]}$. $\rho _{de}/\rho _{tot}=0.73,$ $\rho
_{M}/\rho _{tot}=0.27$, $\rho _{M}=\rho _{VM}+\rho _{DM}$, $\rho _{VM}\sim
\rho _{B},$ $\rho _{DM}/\rho _{tot}=0.23$ and $\rho _{B}/\rho _{tot}=0.04$,
here $\rho _{de}$ is the density of dark energy, $\rho _{tot}$ is the
density of the total energy of the universe $\left( c=1\right) $, $\rho
_{VM} $ is the energy density of visible matter, $\rho _{DM}$ is the energy
density of dark matter, and $\rho _{B}$ is the energy density of visible
baryon matter. According to the cosmological model without singularity$%
^{[5]} $, in the $V-breaking$, the effects of $s-matter$ are equivalent to
that of the so-called dark energy, and $\rho _{v}=\rho _{M}$. According to
this dark-matter model $[3,4]$, because of the symmetry of $F-matter$ and $%
W-matter$, we have 
\begin{eqnarray}
\rho _{M} &=&\rho _{v}=\rho _{vF}+\rho _{vW}=2\rho _{vF},\text{ \ }\rho
_{B}=\rho _{vFB},  \notag \\
\rho _{vF} &=&\rho _{vFB}+\rho _{vFu},\text{ \ }\rho _{vW}=\rho _{vWB}+\rho
_{vWu},  \notag \\
\rho _{vFB} &=&\rho _{vWB},\text{ \ \ }\rho _{vFu}=\rho _{vWu},\text{ \ }%
\rho _{vD}=\rho _{vFu}+\rho _{vW}=\rho _{DM},  \TCItag{9}
\end{eqnarray}%
where $\rho _{v}$ is the total energy density of $v-matter$, $\rho _{vF}$
and $\rho _{vW}$ are respectively the energy density of $v-F-matter$ and the
energy density of $v-W-matter,$ $\rho _{vFB}$ and $\rho _{vWB}$ are
respectively the energy density of $v-F-baryon$ matter $(v-FBM)$ and the
energy density of $v-W-baryon$ matter $(v-WBM)$, $\rho _{vFu}$ is the energy
density of unknown $v-F-matter$ $(v-UFM),$ $\rho _{vWu}$ is the energy
density of $v-W-matter$ $(v-UWM)$ corresponding to $v-UFM,$ and $\rho _{vD}$
is the total energy density of invisible $v-matter$. Here $v-FBM$ is the
given and visible matter which contains given baryon matter, black holes and
neutrinos etc., $F-matter$ contains $v-FBM$ and invisible and unknown $v-UFM$%
. Considering $\rho _{vF}=\rho _{vW}$ because $F-matter$ and $W-matter$ are
symmetric and can transform from one into another when temperature is high
enough, we can determine the ratios of a density to another.

\begin{eqnarray}
\frac{\rho _{vF}}{\rho _{vW}} &=&\frac{0.27/2}{0.27/2}=1=\frac{\rho _{vFB}}{%
\rho _{vW}B}=\frac{\rho _{vFu}}{\rho _{vW}u},  \TCItag{10a} \\
\frac{\rho _{vFB}}{\rho _{v}} &=&\frac{\rho _{vWB}}{\rho _{v}}=\frac{0.04}{%
0.27}=\frac{4}{27},  \TCItag{10b} \\
\frac{\rho _{vFu}}{\rho _{v}} &=&\frac{\rho _{vWu}}{\rho _{v}}=\frac{%
0.27/2-0.04}{0.27}=\frac{9.5}{27},  \TCItag{10c} \\
\frac{\rho _{vD}}{\rho _{v}} &=&\frac{\rho _{vW}+\rho _{vFu}}{\rho _{v}}=%
\frac{23}{27},  \TCItag{10d} \\
\frac{\rho _{vD}}{\rho _{vFB}} &=&\frac{0.27/2+0.095}{0.04}=\frac{23}{4}. 
\TCItag{10e}
\end{eqnarray}

Thus, according to the present model$^{[3]},$ there are three sorts of dark
matter which are $v-UFM,$ $v-WBM$ and $v-UWM.$ Given $v-FBM$ can cluster and
form visible galaxies. $v-WBM$ can cluster and form dark galaxies. $v-UFM$
and $v-UWM$ cannot cluster to form any celestial body, loosely distribute in
space, and their compositions are unknown. $\rho _{vWu}/\rho _{vFu}=\rho
_{vFB}/\rho _{vWB}=1$ is invariant because of the symmetry of $W-matter$ and 
$F-matter$.

According to the cosmological model without singularity$^{[5]},$ there are $%
s-matter$ and $v-matter$ which are symmetric and repulsive each other. In $%
V-breaking$, $v-SU(5)$ breaks into $v-SU(3)\times U(1),$ $v-FBM$ forms the
visible world, $v-WBM,$ $v-UFM$ and $v-UWM$ form the dark-matter world; $%
s-SU(5)$ does not break, all $s-particles$ form $s-SU(5)$ color single
states which loosely distribute in whole space and cause space to expand
with an acceleration in the present stage. This is because there is no
interaction similar to the given electroweak interaction among the $s-SU(5)$
color single states$^{[5]}$.

$\rho _{s}/\rho _{v}$\ is changeable as space expands$^{[5]}$ because $\rho
_{vM}\propto R^{-3},$ $\rho _{v\gamma }\propto R^{-4},$ and $\rho _{s}=\rho
_{sM}\propto R^{-3}.$ Here $\rho _{vM}$ and $\rho _{v\gamma }$ are
respectively the energy density of massive $v-particles$ and the energy
density of massless $v-particles$, $\rho _{s}$ is the energy density of $%
s-SU(5)$ color single states. The masses of all color single states are not
zero, hence $\rho _{s}\propto R^{-3}.$

\section{The interaction of F-scalar fields and W-scalar fields}

The Ref. $[7]$ given the interaction of $F-scalar$ fields and $W-scalar$
fields

\begin{equation}
V=-\frac{2A}{225}Tr\Phi _{f}^{2}Tr\Phi _{w}^{2},  \tag{11}
\end{equation}%
where $\Phi _{f}$ and $\Phi _{w}$\ \ are\ the $\underline{24}$
representation of $SU(5)$ group. The breaking component is 
\begin{equation}
\Phi _{i}=Diagonal\left( 1,1,1,-\frac{3}{2},-\frac{3}{2}\right) \left(
\sigma _{i}+\varphi _{i}\right) ,  \tag{12}
\end{equation}%
where the subscript $i=f,$ $w.$ From $\left( 11\right) -\left( 12\right) $
we obtain 
\begin{equation}
V_{fw}=-A\left( 2\sigma _{f}\sigma _{w}\varphi _{f}\varphi _{w}+\sigma
_{f}\varphi _{f}\varphi _{w}^{2}+\sigma _{w}\varphi _{w}\varphi _{f}^{2}+%
\frac{1}{2}\varphi _{f}^{2}\varphi _{w}^{2}\right) .  \tag{13}
\end{equation}%
$\mid \sigma _{w}\mid =\mid \sigma _{f}\mid $ because of the symmetry of $%
s-matter$ and $f-matter$. Both $\sigma _{i}$ and $m\left( \varphi
_{i}\right) $ are functions of temperature $T$. When $T\geq T_{cr},$ $\sigma
_{i}=m\left( \varphi _{i}\right) =0$. Consequently $f-particles$ and $%
w-particles$ can easily transform from one to another so that $\rho
_{F}=\rho _{W}.$ When $T\sim 0,$ both $\mid \sigma _{i}\mid $and $m\left(
\varphi _{i}\right) $ are large enough. Consequently interaction between $%
f-particles$ and $w-particles$ by the scalar bosons may be ignored. Thus
there is only the gravitation between $f-matter$ and $w-matter$ when
temperature is low$.$

There are the couplings of fermions (and gauge particles) and scalar bosons$%
^{[7]}.$ Hence there are the interactions of $f-fermions$ and $w-fermions$
shown in figures 1-3 and the interactions of $f-gauge$ bosons and $w-gauge$
bosons via the scalar bosons $\varphi _{f}$ and $\varphi _{w}.$ In the
figures the dotted lines with arrows denote $W-fermion$ field $\psi _{w}$,
the dotted lines without arrow denote $W-scalar$ field $\varphi _{w},$ the
lines with arrows denote $F-fermion$ field $\psi _{f}$, the lines without
arrow denote $F-scalar$ field $\varphi _{f},$ $M^{2}=-2A\sigma _{f}\sigma
_{w}$, $R_{f}=-A\sigma _{f}$, $R_{w}=-A\sigma _{w}$ and $S=-A/2.$

It can be seen from figure 1 and $\left( 13\right) $ that when $-A\sigma
_{f}\sigma _{w}>0$\ and $k^{2}-m^{2}<0$ or $-A\sigma _{f}\sigma _{w}<0$\ and 
$k^{2}-m^{2}>0,$\ $f-fermions$ and $w-fermions$\ are repulsive each other;
when $-A\sigma _{f}\sigma _{w}>0$\ and $k^{2}-m^{2}>0$ or $-A\sigma
_{f}\sigma _{w}<0$\ and $k^{2}-m^{2}<0,$\ $f-fermions$ and $w-fermions$\ are
attractive each other.\FRAME{ftbpFU}{3.8303in}{3.205in}{0pt}{\Qcb{$f+%
\overline{f}\longrightarrow w+\overline{w}$ realized by the tree digram.}}{%
\Qlb{Fig.1}}{cd-1.ps}{\special{language "Scientific Word";type
"GRAPHIC";display "USEDEF";valid_file "F";width 3.8303in;height
3.205in;depth 0pt;original-width 7.8646in;original-height 11.3481in;cropleft
"0.0643";croptop "0.8844";cropright "0.9356";cropbottom "0.2744";filename
'H: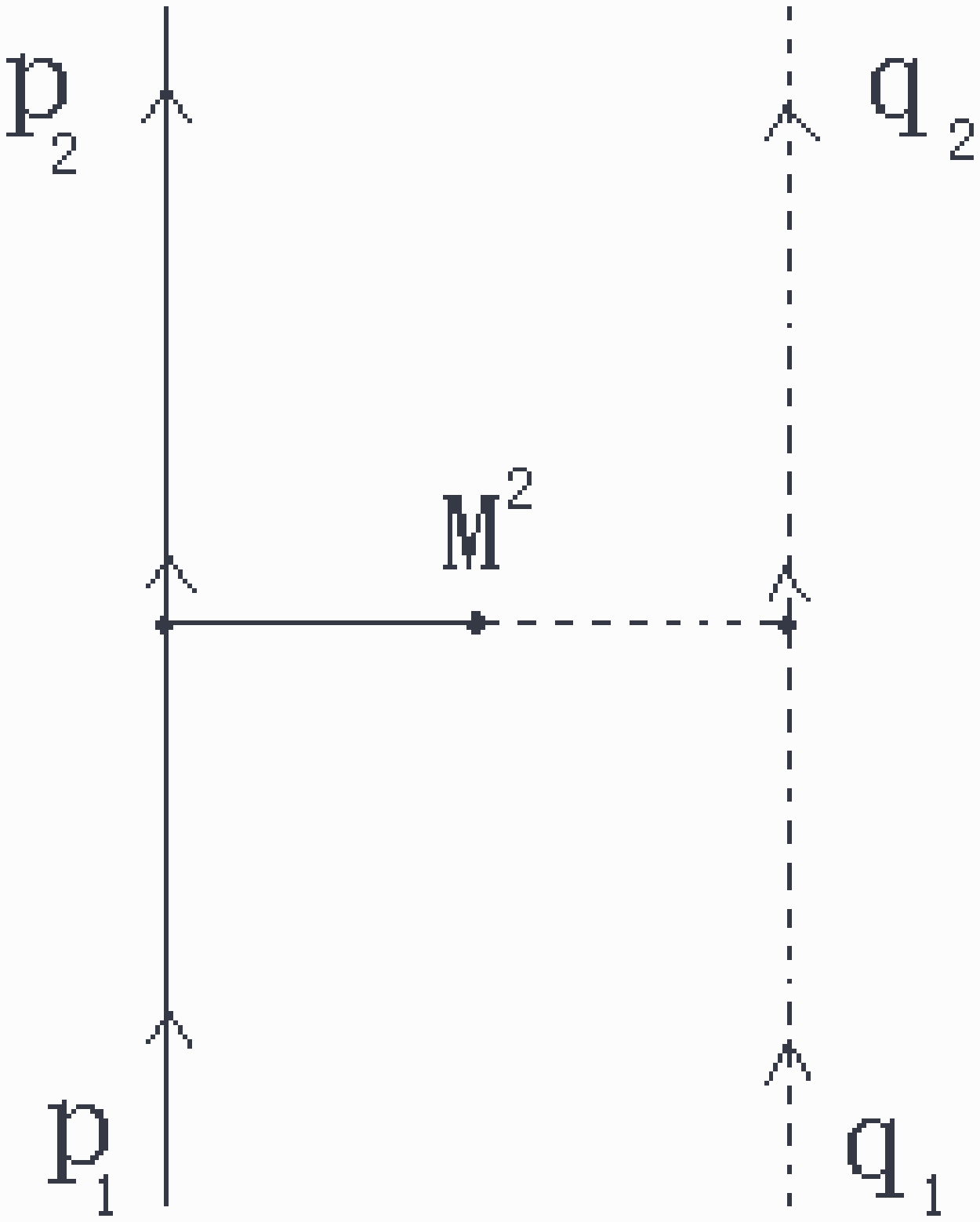';file-properties "XNPEU";}}\FRAME{ftbpFU}{5.4691in}{%
3.9531in}{0pt}{\Qcb{$f+\overline{f}\longrightarrow w+\overline{w}$ realized
by the two one-loop digrams.}}{\Qlb{Fig.2}}{cd-2.ps}{\special{language
"Scientific Word";type "GRAPHIC";display "USEDEF";valid_file "F";width
5.4691in;height 3.9531in;depth 0pt;original-width 7.8646in;original-height
11.3481in;cropleft "0.0978";croptop "0.9512";cropright "0.9021";cropbottom
"0.5273";filename 'H: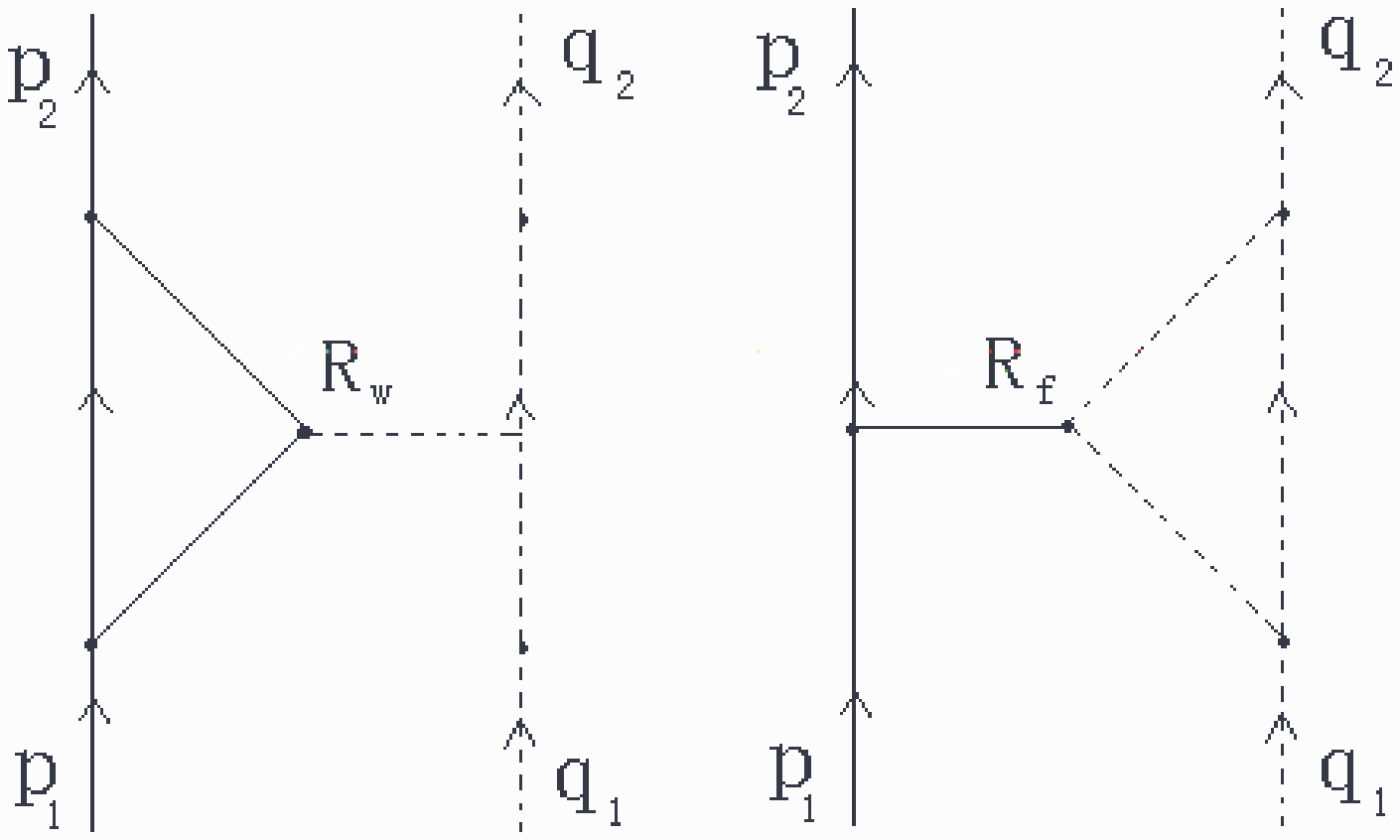';file-properties "XNPEU";}}

\FRAME{ftbpFU}{3.2465in}{3.1254in}{0pt}{\Qcb{$f+\overline{f}\longrightarrow
w+\overline{w}$ realized by the two-loop diagram.}}{\Qlb{Fig.3}}{cd-3.ps}{%
\special{language "Scientific Word";type "GRAPHIC";maintain-aspect-ratio
TRUE;display "USEDEF";valid_file "F";width 3.2465in;height 3.1254in;depth
0pt;original-width 7.8646in;original-height 11.3481in;cropleft "0";croptop
"0.9330";cropright "1";cropbottom "0.2660";filename
'H: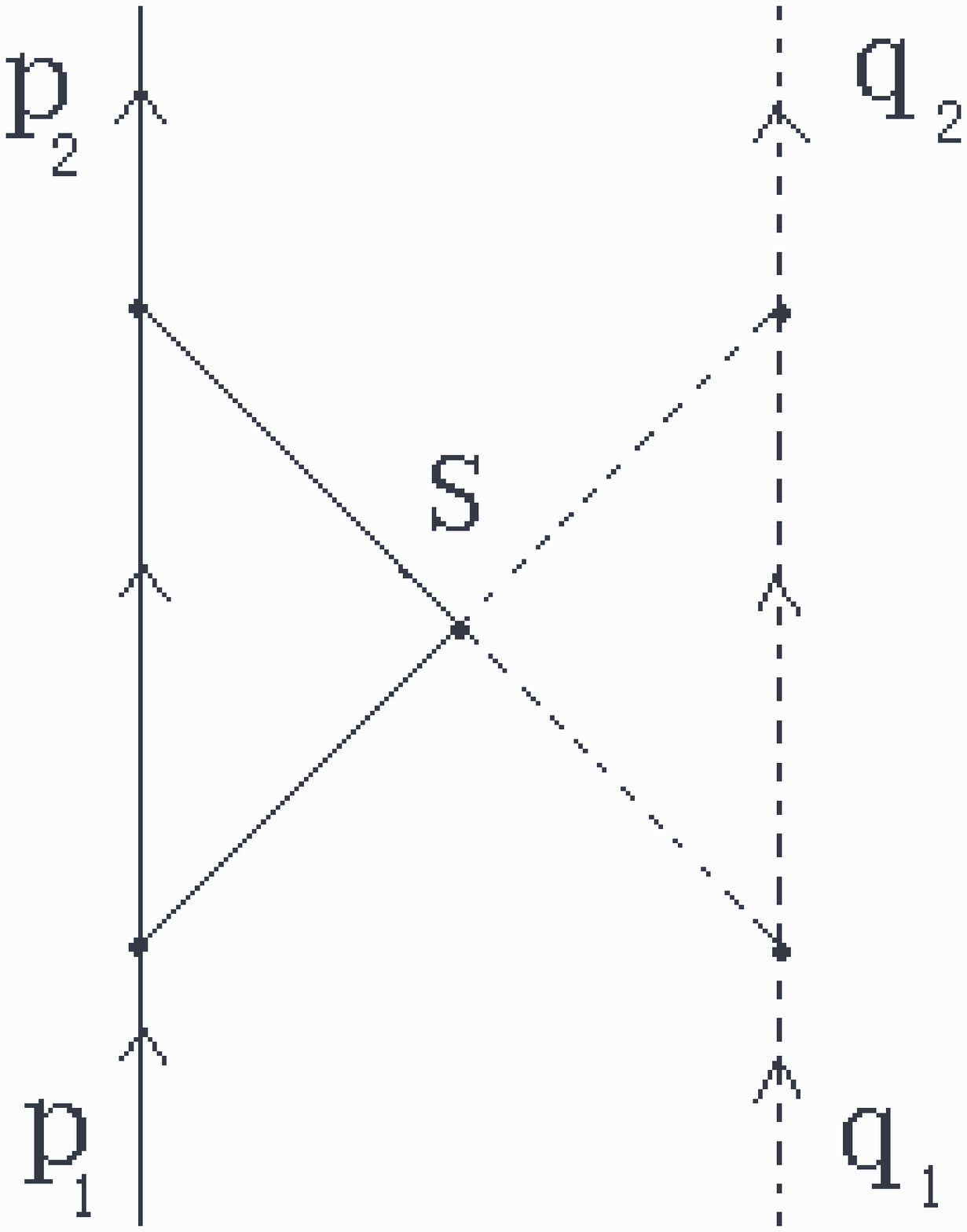';file-properties "XNPEU";}}

\section{Features and observation of dark matter in present model}

According to the present model$^{[3,4]}$, $v-UFM$ and $v-UWM$ cannot form
cluster, loosely distribute in space and have positive gravitational masses,
hence both should be identified as cold dark matter. From $\left( 10\right) $
we have 
\begin{eqnarray}
\frac{\rho _{vFu}+\rho _{vWu}}{\rho _{v}} &=&\frac{0.095\times 2}{0.27}=%
\frac{19}{27},  \notag \\
\frac{\rho _{vFu}+\rho _{vWu}}{\rho _{vD}} &=&\frac{0.095\times 2}{0.23}=%
\frac{19}{23}.  \TCItag{14}
\end{eqnarray}%
It is seen that the present model cannot differ from the cold dark matter
model by $v-UFM$ and $v-UWM.$ $W-baryon$ matter $v-WBM$ can form clusters of
dark matter and dark galaxies$^{[3,4]}$. There possibly are such celestial
bodies which are composed of mixture of $F-matter$ and $W-matter$ because
both have positive gravitational masses. This is a decisive predict of the
present model and mirror dark matter model. Dark celestial bodies flying to
the earth are possibly detected by probing gravity-meters of clustering dark
matter$^{[4]}$.

It can be seen from $\left( 13\right) $ and figures $1-3$ that the energy of 
$F-matter$ can transform into the energy of $W-matter$ by such a process in
which the reaction energy is high enough.

\section{Conclusions}

According to the cosmological model without singularity, there are $s-matter$
and $v-matter$ which are symmetric and have oppose gravitational masses. In $%
V-breaking$ $s-matter$ is similar to dark energy to cause expansion of the
universe with an acceleration now, and $v-matter$ is composed of $%
v-F-matterv $ and $v-W-matterv$ which are symmetric and have the same
gravitational masses$.$ The ratio of $s-matter$ to $v-matter$ is changeable.

Based on the cosmological model, we confirm that big bang nucleosynthesis is
not spoiled by that the average energy density of $W-matter$ (mirror matter)
is equal to that of $F-matter$ (ordinary matter).

According to the present model, there are three sorts of dark matter which
are $v-W-baryon$ matter ($4/27)$ and unknown $v-F-matterv$ ($9.5/27)$ and $%
v-W-matter$ ($9.5/27).$ Given $v-F-bayon$ matter ($4/27)$ can cluster to
form the visible galaxies. $V-W-bayon$ matter can cluster to form dark
celestial bodies and dark galaxies. Unknown $v-F-matterv$ and $v-W-matter$
cannot cluster to form any celestial body, loosely distribute in space, are
equivalent to cold matter, and their compositions are unknown. The number in
a bracket is the ratio of the density of a sort of matter to total density
of $v-matter$.

The decisive predict is that there are dark celestial bodies and dark
galaxies. The energy of $F-matter$ can transform into the energy of $%
W-matter $ by such a process in which the reaction energy is high enough.

\textbf{Acknowledgement}

I am very grateful to professor Zhao Zhan-yue and professor Wu Zhao-yan for
their helpful discussions and best support.

\end{document}